# X-ray Diffraction and Equation of State of the C-S-H Room-Temperature Superconductor

Anmol Lamichhane,[1,*] Ravhi Kumar,[1] Muhtar Ahart,[1] Nilesh Salke,[1] Nathan Dasenbrock-Gammon,[2] Elliot Snider,[3] Yue Meng,[4] Barbara Lavina,[5,6] Stella Chariton,[6] Vitali B. Prakapenka,[6] Maddury Somayazulu,[4] Ranga P. Dias,[2,3] and Russell J. Hemley[1,7,*]

[1]*Department of Physics, University of Illinois at Chicago, Chicago IL 60607*
[2]*Department of Physics and Astronomy, University of Rochester, Rochester, NY 14623*
[3]*Department of Mechanical Engineering, University of Rochester, Rochester, NY 14623*
[4]*HPCAT, X-ray Science Division, Argonne National Laboratory, Argonne IL 60439*
[5]*Center for Advanced Radiation Sources, University of Chicago, Chicago, IL 60439*
[6]*X-ray Science Division, Argonne National Laboratory, Argonne IL 60439*
[7]*Department of Chemistry, University of Illinois at Chicago, Chicago IL 60607*

X-ray diffraction indicates that the structure of the recently discovered carbonaceous sulfur hydride (C-S-H) room temperature superconductor is derived from previously established van der Waals compounds found in the $H_2S$-$H_2$ and $CH_4$-$H_2$ systems. Crystals of the superconducting phase were produced by a photochemical synthesis technique leading to the superconducting critical temperature $T_c$ of 288 K at 267 GPa. Single-crystal x-ray diffraction patterns measured from 124 to 178 GPa, within the pressure range of the superconducting phase, give an orthorhombic structure derived from the $Al_2Cu$-type determined for $(H_2S)_2H_2$ and $(CH_4)_2H_2$ that differs from those predicted and observed for the S-H system to these pressures. The formation and stability of the C-S-H compound can be understood in terms of the close similarity in effective volumes of the $H_2S$ and $CH_4$ components, and denser carbon-bearing S-H phases may form at higher pressures. The results are crucial for understanding the very high superconducting $T_c$ found in the C-S-H system at megabar pressures.

* alamic2@uic.edu; rhemley@uic.edu

**Introduction**

The seminal prediction that chemical doping could lower the transition pressure of atomicmetallic and superconducting hydrogen[1] inspired the exploration of the high-pressure behavior of simple molecules mixed with hydrogen under pressure.[2-4] Later theoretical calculations predicted new classes of novel hydrogen-rich solids exhibiting very high $T_c$ behavior when such molecular mixtures are subjected to megabar pressures.[5-9] Resulting experimental efforts in the last few years have led to the discovery of $T_c$ above 203 K in $H_3S$,[10] up to 260 K LaH$_{10}$,[11-14] and later, similar values for Y-H hydrides.[15-17] These studies include other low-Z materials such as possible carbon-bearing high $T_c$ hydrides in view of their possible stability at lower, and even ambient, pressures[18] as well as predicted incorporation of other components within these structures to form very high $T_c$ chemically doped[19-21] or stoichiometric perovskite-like structures.[22-24]

Recently, high temperature superconductivity with a critical temperature of 288 K at 267 GPa has been reported in the C-S-H system.[25] Information on the low-pressure precursor phases were obtained from Raman spectroscopy together with theoretical calculations. The Raman data revealed the formation of C-H bonding as well as phase transitions at 15 GPa and 37 GPa, but spectra were not measurable above 60 GPa when the samples became opaque. The formation of C-H bonding prior to the darkening of the phase that becomes superconducting, leading to the expectation of carbon uptake, persists into the superconducting phase.[25] However, no diffraction or direct information on the structure of the material in the superconducting state was reported. Subsequent electron-phonon calculations indicated that incorporation of a small amount (*i.e.,* 4 atom %) in the structure could produce the very high $T_c$ by hole doping of $H_3S$.[20,21]

Here we report single crystal x-ray diffraction data and the *P-V* equation of state (EOS) of the very high $T_c$ superconducting material reported in Ref. [25]. X-ray diffraction indicates that the structure is derived from previously established van der Waals compounds found in the $CH_4$-$H_2$ (Ref. [3]) and $H_2S$-$H_2$ (Ref. [4]) systems but has a lower symmetry over the measured range. The structure differs in detail from those found and predicted for phases in the S-H system. The results provide a mechanism for incorporation of carbon in molecular hydrogen sulfide structures starting at low pressures and are consistent with the hole-doping model for the room-temperature superconductivity proposed for the C-S-H system at megabar pressures.

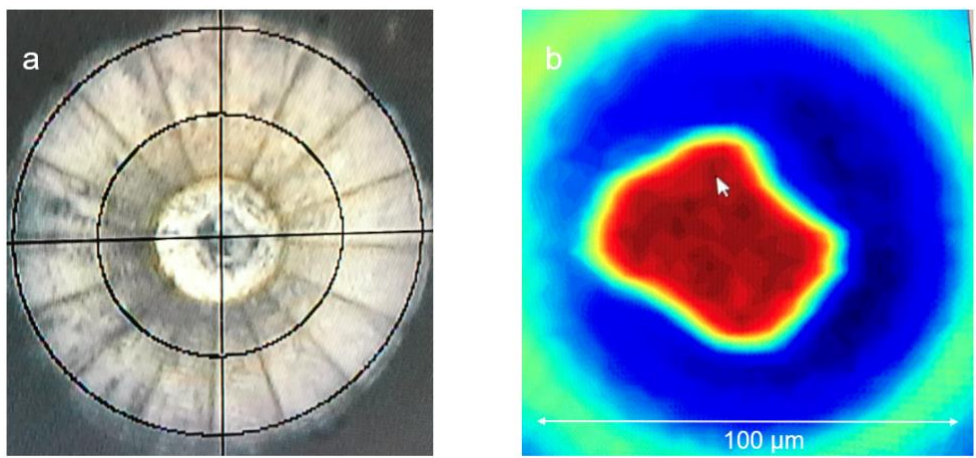

*Figure 1. Images of the sample in a diamond anvil cell at 124(2) GPa (run 22 from Ref. [25]). (a) Combined transmission/reflection photomicrograph showing the optically dark superconducting C-S-H sample at the center of the cross-hairs within the sample chamber (grey region); the white area is the highly reflecting Re gasket on the 100-μm culet of the beveled diamond anvil. (b) X-ray transmission image of a magnified view of the sample chamber (red) showing the C-S-H sample and the diamond culet (blue).*

**Results**

Images of the sample at 124 GPa prepared by the procedures described above are shown in Fig. 1. The x-ray and optical imaging show the size of sample to be around 40 μm. Powder x-ray diffraction mapping of the sample in the region surrounding the optically dark area matches high-pressure x-ray patterns reported for disordered H carbon[26] and, at the sample edges, Re and Re-H gasket material, which was also used for pressure calibration. Numerous experiments show that the optically dark region is the superconducting sample at this pressure.[25] X-ray diffraction reveals the dark region to be largely single-crystalline. Examples of diffraction images are shown in Fig. 2.

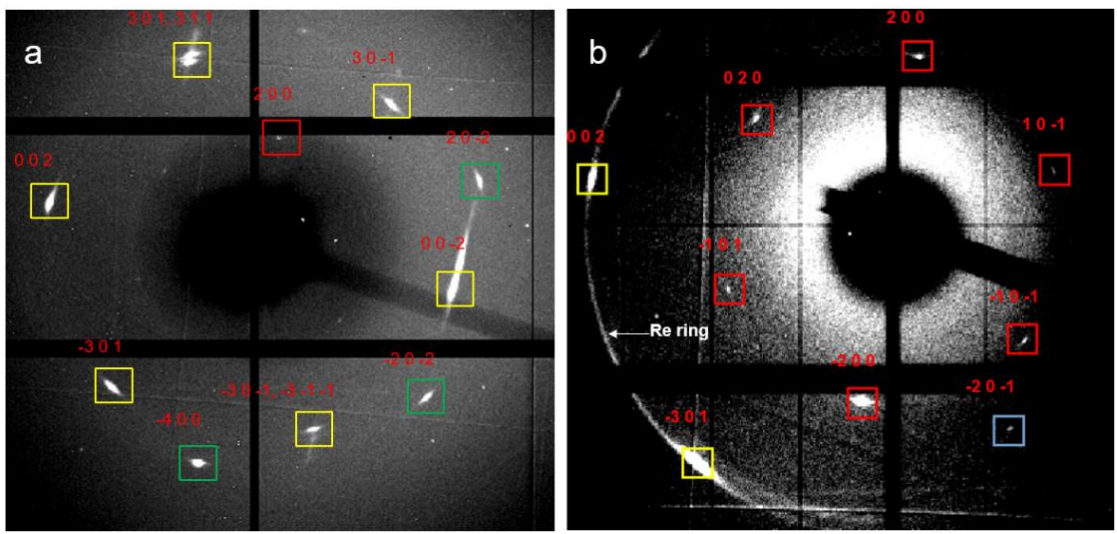

*Figure 2. Single-crystal diffraction images of the C-S-H sample at 124(2) GPa. (a) Image obtained at GSECARS showing symmetry equivalent reflections corresponding to (200), (002)/(102) and (022) planes labeled with red, yellow and green boxes, respectively. The streak near the 00-2 reflection is due to disorder. (b) Image obtained at HPCAT showing symmetry equivalent reflections corresponding to (200)/(020), (201) and (002) planes labeled with red, blue and yellow boxes, respectively. The ring appearing on the left side of the image is polycrystalline diffraction from the Re gasket. The black circles in the center of both images are shadows of the x-ray beam stop, and the black crosses arise from the detector. Images as a function of pressure are provided in the Supplementary Material.*

For each of the images, we obtained $d$-spacings by averaging the values of symmetry equivalent peaks (Fig. 3). The prominent reflections could be identified as belonging to planes (110), (002), (121) and (112) (GSECARS) and (110), (002), and (111) (HPCAT) of a cell that is close to tetragonal. However, the (110) reflections for both gave two different populations of $d$-spacings indicating a weak orthorhombic distortion that can be interpreted as a $110_t$ splitting into $200/020_o$ with twice the unit cell volume. Moreover, the observation of the $111_t (= 201_o)$ reflections at all pressures rules out the *I4/mcm* and *Cccm* space groups that had been observed or predicted for structures in the binary systems of $CH_4$-$H_2$ and $H_2S$-$H_2$.[3,4,7,27-30] X-ray diffraction mapping of the sample showed no symmetry lowering due to possible pressure gradients. The lattice parameters obtained, for example at 124(2) GPa, give $a$ = 7.76(2) Å, $b$ = 7.70(2) Å, $c$ = 4.48(2) Å and $V$ = 16.73(3) Å$^3$/ S, with a smooth evolution with pressure (Fig. 4). The data indicate a nearly constant $c/a$ (and $c/b$), both 0.82(2), with respect to a tetragonal cell over the measured pressure range (for example, $\sqrt{2}(c/a)_o \to (c/a)_t$ and $\sqrt{2}(c/b)_o \to (c/b)_t$), which is essentially the same as that found for *I4/mcm* $CH_4$-$H_2$ [3] and $H_2S$-$H_2$ [4] near 5 GPa.

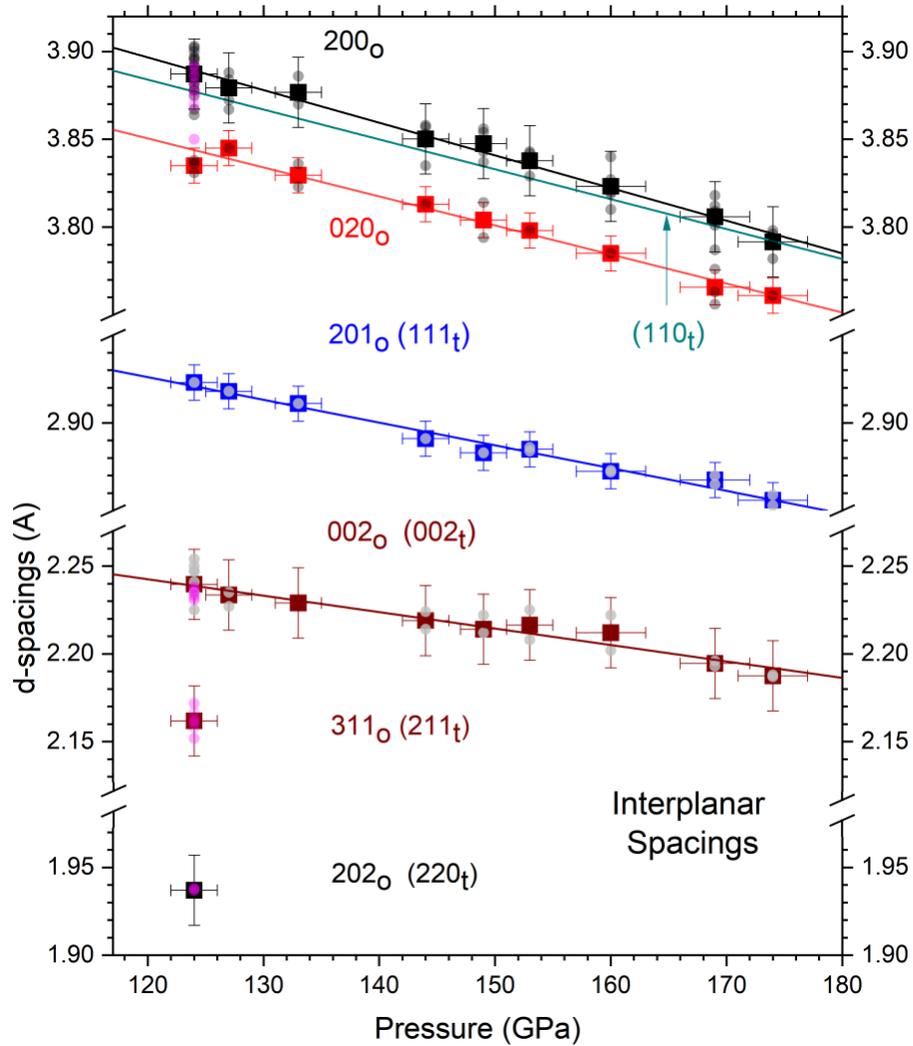

*Figure 3*. Pressure dependence of d-spacings determined from integration of single-crystal reflections. The solid squares correspond to the average of the d-spacing values measured by single-crystal diffraction, which are shown by transparent grey circles (HPCAT) and transparent pink circles (GSECARS). The solid lines are fits through the data points assuming an orthorhombic structure, i.e., $(hkl)_o$; the fit of the data assuming a tetragonal structure is also shown for $110_t$. In this setting, $201_o$ and $002_o$ correspond to $111_t$ and $002_t$, respectively. The uncertainties in the d-spacings for the averaged $200_o$ data points correspond to $2\sigma$ in the measured distribution of the d-spacings, whereas those for $020_o$ and $201_o$ are estimated from the observed spread in the data.

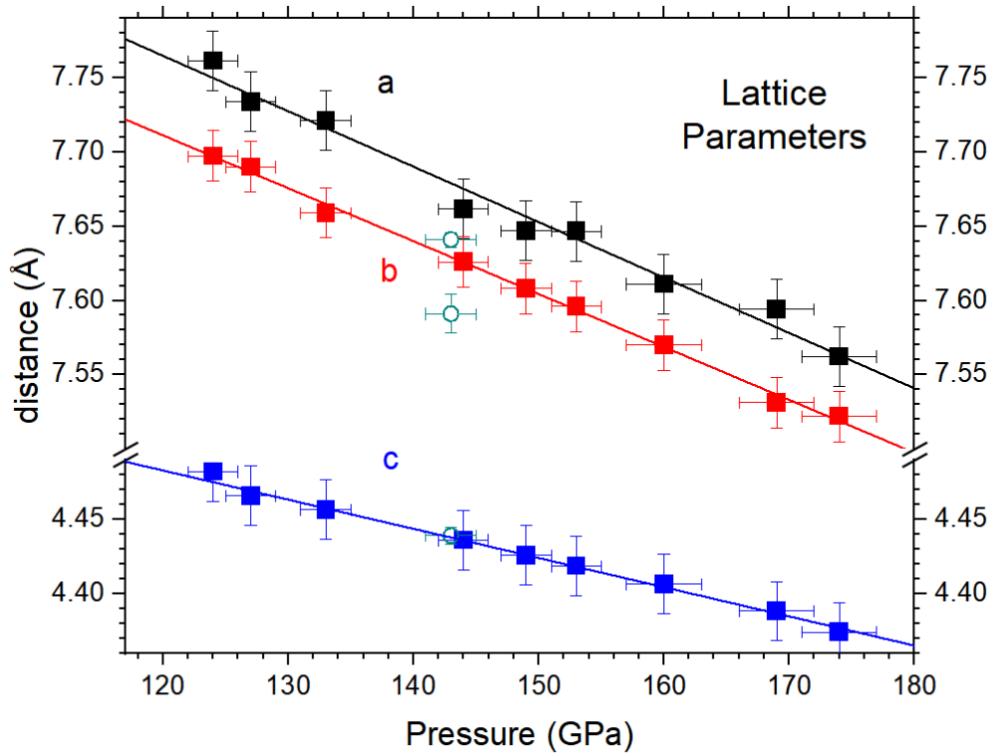

*Figure 4.* Pressure dependence of the lattice parameters determined from d-spacing values obtained in the orthorhombic crystal system. Solid squares are lattice parameters from the average of the d-spacing data at a given pressure. Open circles are lattice parameters of a structure reported after laser heating a C-S-H sample at 143 GPa.[31]

Figure 5 shows the *P-V* relations obtained from diffraction data as a function of pressure from 124 to 178 GPa. We also compare the measured volumes reported for the C-S-H material by Bykova et al.[31] and those reported for S-H phases.[27-30,32,33] The difference between the EOS obtained here for C-S-H and that reported for $H_3S$ (Ref. [32]) is about 1.4 Å$^3$ / S, C across the indicated pressure range. The room-temperature compression data reported for a C-S-H sample of Bykova et al.[31] and for $(H_2S)_2H_2$ by Guigue et al.[27] differ from that found here. The molecular volumes of *Cccm* $(H_2S)_2H_2$ identified by Pace et. al.[29] are close to those measured here for the C-S-H sample. The nearly constant difference between the volumes for C-S-H and $H_3S$ over the measured pressure range contrasts with the trend reported for room-temperature compression of a C-S-H sample by Bykova et al.[31] On the other hand, the *P-V* point for that sample after laser heating to form an orthorhombic structure is closer to the EOS measured here.

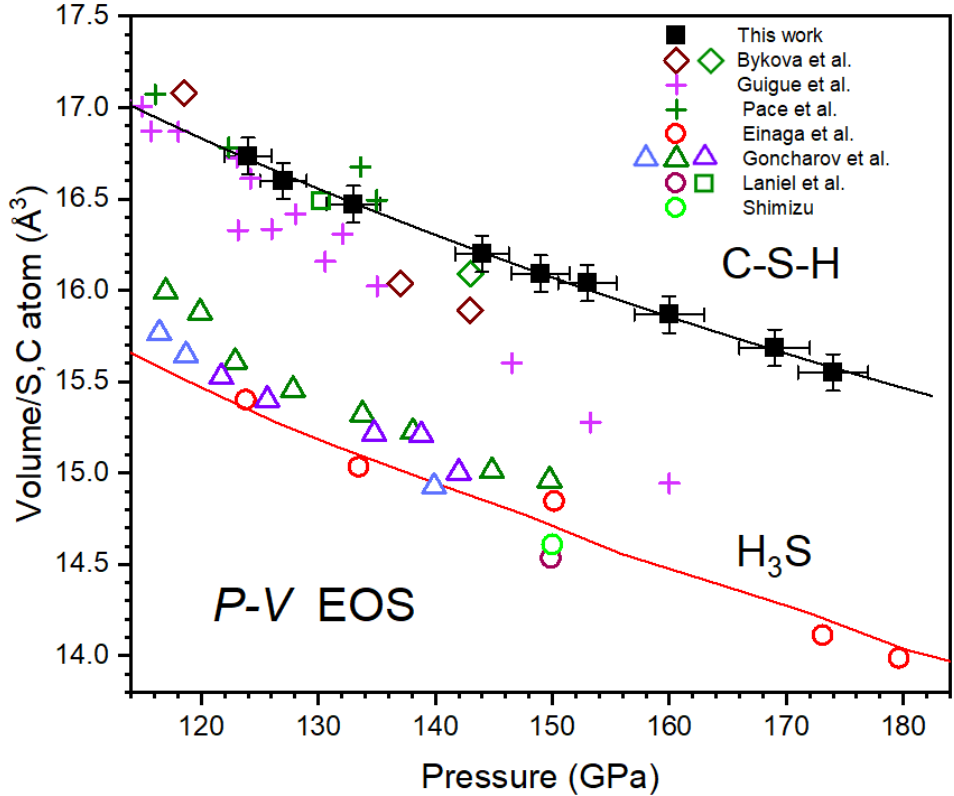

*Figure 5.* Pressure-volume (P-V) relations measured for C-S-H compared to data from experiments on $H_3S$ composition samples. Filled squares are from the orthorhombic structure obtained in the present work, and the black line is an EOS fit. The volume and pressure uncertainties were determined from analysis of the d-spacings (fig S1) and the Re x-ray diffraction peaks. Open diamonds correspond to the C-S-H data of Bykova et al.,[31] where the brown diamond is that of the tetragonal structure (identified as I4/mcm) after lower pressure synthesis and room-temperature compression, and the green diamond that of the orthorhombic structure (identified as Pnma) produced after laser heating of that sample. Open circles correspond to the proposed structures of $H_3S$ composition by Einaga et al.[32] (two highest data points are for $D_3S$), Shimizu,[33] and Laniel et al.[30]; open triangles, Goncharov et. al.;[28] crosses, Cccm phase of Guigue et al.[27] and Pace et al.[29] The red line is the EOS fit from Einaga et al.[32] The green diamond corresponds to the I4/mcm structure reported by Laniel et al.[30] to be hydrogen sulfide after laser heating at lower pressure, but the sample may contain carbon from the paraffin used as a hydrogen source in the synthesis.

The *P-V* EOS relations for phases in the C-S-H system over an extended pressure range are presented in Fig. 6. The results show that our measurements of the compression of the C-S-H superconductor can be viewed as an extension of the EOS of $(H_2S)_2H_2$ and $(CH_4)_2H_2$, which have essentially identical volumes in their *I4/mcm* structures at synthesis pressures near 5 GPa. The atomic ratios for such $Al_2Cu$-related structures range from 1.13:1 to 1.61:1.[3] Experimentally constrained EOS curves for $H_2S$,[34] $CH_4$,[35] C (diamond),[37] and $H_2$[36] phases are also shown in Fig. 6. $CH_4$ and $H_2S$ have similar molecular volumes over a wide range of pressures, consistent with the similarity in molar volumes of $(CH_4)_2H_2$ and $(H_2S)_2H_2$ van der Waals compounds. The EOS measurements and analyses indicate that the effective molecular volumes of $H_2S$ and $CH_4$ are comparable over a broad range of pressures (Fig. 6). For example, at 124 GPa, effective volumes are both around 14 $Å^3$ / molecule. The $H_2$ volume decreases from roughly 13 $Å^3$ at the room-temperature freezing pressure of 5.4 GPa to 4.3 $Å^3$ at the pressures studied here.[36] We also compare the volumes of the $CH_4 + H_3S$ molecular[27,32,40] and the $C + S + 7H$ elemental[36,38,41] assemblages. The volume for $CSH_7$ ($H_3S$-$CH_4$ perovskite) is lower than that of $H_3S$, consistent with the smaller volume of methane, and is metastable with respect to decomposition to $CH_4$ and $H_3S$ as described in Ref. [22]

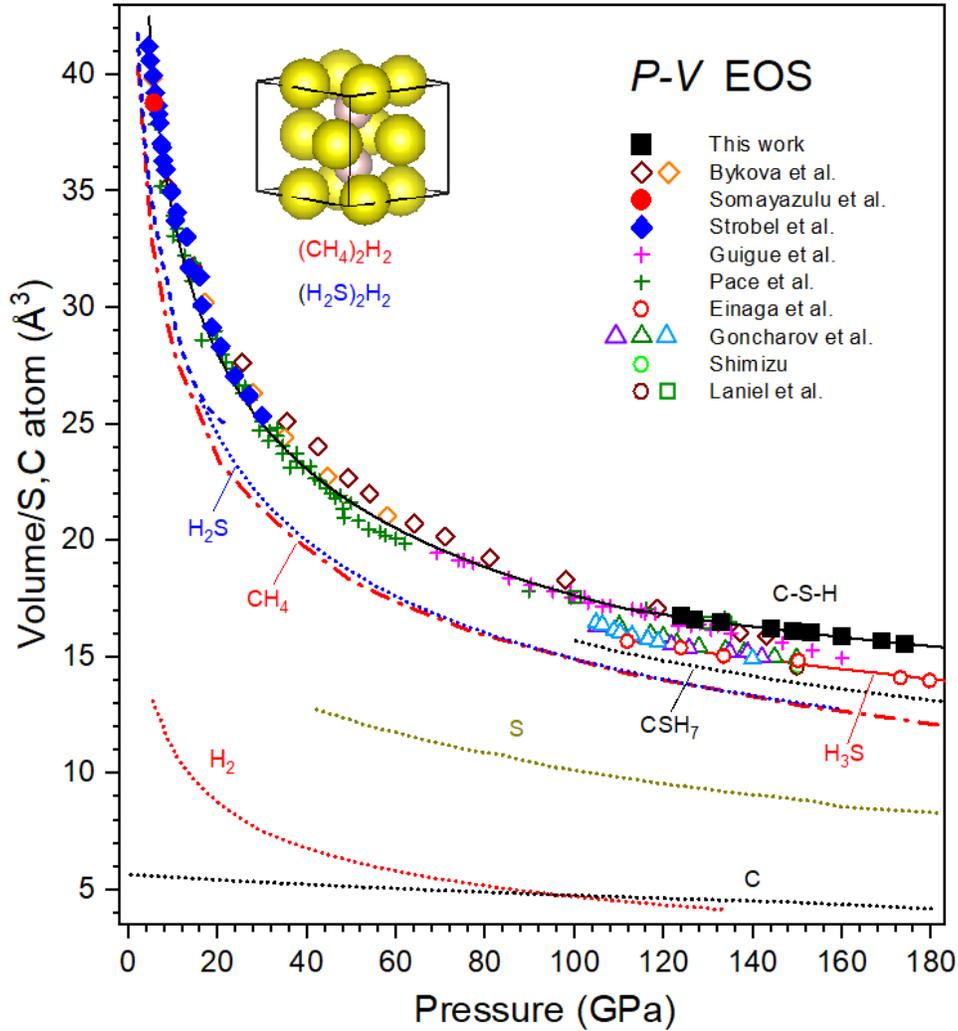

*Figure 6. Extended P-V EOS comparison of phases and phase components in the C-S-H system. The references for data points for the C-S-H and S-H compounds are given in the caption to Fig. 5. The data point for $CH_4$-$H_2$ is from Somayazulu et al.[3] The black line is a continuation to low pressures of the EOS fit of the C-S-H points. The yellow spheres in the structure model drawing correspond to $H_2S$ and $CH_4$, whereas the pink spheres represent rotationally disordered (or J = 0) $H_2$ molecules. The hydrogen positions are not determined in these x-ray diffraction experiments. Hydrogen bonding between $H_2S$, including pressure-induced symmetric hydrogen bonding under pressure, is expected.[4] The EOS of $H_2S$,[34] $CH_4$,[35] $H_2$ ($Å^3$/molecule),[36] C (diamond),[37] and S[38] phases are shown. For $H_2S$, the EOS for the data reported in Ref. [34] as well as a fit to data below 20 GPa and extrapolated to higher pressures are shown (see also Ref. [30]). The $CH_4$ curve is a theoretical EOS[35] that agrees with experimental data over a wide range of compression,[39,40] and the diamond result[37] is consistent with earlier data.[41] The calculated static lattice EOS for $CSH_7$ perovskite[22] is also compared. The calculated P-V curves for the competitive low-enthalpy structures considered in that study are within 0.5% of each other.*

**Discussion**

We now comment on the implications of the above x-ray diffraction and EOS data for the structure, chemical interactions, and very high-$T_c$ superconductivity documented in the recent studies of the C-S-H system under pressure. The structure is derived from the high-symmetry $I4/mcm$ structure observed for the two van der Waals compounds. The stoichiometry may thus be written $[(CH_4)_2H_2]_x[(H_2S)_2H_2]_y$; although $x$ and $y$ cannot be determined from the x-ray diffraction data, $y \gg x$ is suggested by the similarity in initial $T_c$ of C-S-H and $H_3S$. On the other hand, the structure of the material over the pressure range studied has a lower symmetry than that theoretically predicted or found in the pure $H_2S$ or $CH_4$ compounds. The similarity in size between $CH_4$ and $H_2S$ allows for facile mixing of molecules to form pseudo-ternary $H_2S$-$CH_4$-$H_2$ van der Waals compounds during photochemically-induced synthesis of the precursor phase (or phases) to the superconductor. The constant volume difference of about 1.4 Å$^3$ / S,C for the C-S-H and $H_3S$ EOS suggest that $H_2$ in the parent $I4/mcm$ structure is retained in the structure at these pressures. The difference in volume between $H_3S$ and C-S-H could also arise from the incorporation of additional C or H in the S-H structure. It is possible that at higher pressures, residual molecular components in the structure such as $H_2$ undergo pressure-induced dissociation and the structure collapses to form a denser framework with an EOS similar to that shown for $Im\bar{3}m$ $H_3S$.

Density-functional theory calculations have predicted the stability of hydride perovskite or intercalation $H_3S$-$CH_4$ compounds with a ground state orthorhombic ($Pnma$) and cubic ($I\bar{4}3m$) structures that are nearly degenerate at these pressures.[22,23] The predicted orthorhombic phase differs from that observed experimentally and has a higher density (Fig. 6). However, much less incorporation of $CH_4$ in the $H_3S$ than previously considered is possible in such perovskite-like structures. This could give rise to stoichiometric $H_3S(CH_4)_n$ compounds corresponding to Magneli phases; alternatively, alloys having a broad range of compositions and hydrogen content are possible, both of which would provide another mechanism for the incorporation of carbon in the hydrogen sulfide. Density functional theory calculations assuming BCS/Eliashberg model electron-phonon coupling and using the virtual crystal approximation (VCA) predict that the high $T_c$ observed in the C-S-H structure can arise from hole doping of carbon in $H_3S$.[20,21] The parent cubic ($Im\bar{3}m$) phase of $H_3S$ was assumed as a model structure in these calculations. We point out that the apparent increase in $T_c$ near 200 GPa suggested a transition at that pressure.[25] However, given the scatter in the data, the results are also consistent with a continuous increase in $T_c$ with pressure from 120 GPa to 267 GPa, a result that is also consistent with the theoretical calculations.[20,21]

Bykova et al.[31] reported that the tetragonal $I4/mcm$ structure for one of their C-S-H samples transformed to an orthorhombic structure upon laser heating near 143 GPa. As described above, the volume and lattice parameters are close to the trend found here for the C-S-H sample synthesized photochemically at low pressures. Synthesis of high-pressure superconducting phases in the S-H system in general is path dependent,[10] and this is likely to be especially true for synthesis of high-quality crystals of the C-S-H superconductor, which is facilitated by laser irradiation at specific pressures (see Ref. [25]). We also point out that the hydrogen sulfide samples studied by Laniel et al.[30] may contain carbon from the paraffin used as the hydrogen source in their experiments. Variable carbon content in the phases is likely to arise from differences in sample preparation. Although nearly 1:1 molar ratios of C and S were used in the synthesis of the present samples,[17] the composition of the $CH_4$-$H_2S$-$H_2$ van der Waals compound or alloy that is a precursor to the superconducting phase is not known. Indeed, variations in the alloying in the initial $CH_4$- $H_2S$-$H_2$ mixture are likely to give rise to slight differences in the reported low-pressure Raman spectra.[25,31] The methods used in the preparation of the samples studied here produced the very high $T_c$ superconductor; although not directly measured, at the maximum pressures of the diffraction study the $T_c$ is estimated to be 180 K.[25]

In conclusion, single-crystal x-ray diffraction indicates that the structure of the high $T_c$ superconductor formed in the C-S-H system up to 178 GPa is derived from structures found at lower pressure in the $H_2S$-$H_2$[4] and $CH_4$-$H_2$[3] van der Waals compounds. The mechanism of formation and stability of the material can be understood in terms of the pressure dependence of the effective volumes of $CH_4$, $H_2S$, $H_3S$, and $H_2$. Although more detailed structural information is possible from dedicated single-crystal diffraction with broad reciprocal space coverage, the present measurements on C-S-H samples synthesized for observations of superconductivity establish the overall unit cell symmetry and compressional behavior of the material. As such, the results provide crucial constraints for theoretical studies of the detailed mechanism of the observed room-temperature superconductivity. The incorporation of carbon in hydrogen sulfide is broadly consistent with hole-doping of $H_3S$ by carbon predicted theoretically, but further studies are required for definitive tests. The bonding, coordination, and amount of carbon as a dopant could be determined by x-ray Raman measurements.[42] At higher pressures, the material may evolve toward denser framework structures, ideally examined using combined measurements of transport and structural properties. Alternative synthesis routes and chemical stabilization of the superconductor at lower pressures should also be explored.

**Experimental Methods**

The samples were prepared using the diamond-anvil cell (DAC) techniques described previously (runs 19-22 of Ref. [25]). Raman spectra collected after photochemically assisted synthesis have features found in earlier studies of $CH_4$-$H_2$ and $(H_2S)_2H_2$ (see Supplementary Material). X-ray diffraction patterns were collected at the IDB station of HPCAT (Sector 16, Advanced Photon Source) using an incident wavelength of 0.4066 Å, and at the IDD station of GSECARS (Sector 13, APS) with an incident wavelength of 0.2952 Å. The principal results reported here are for run 22. Because the DACs used were designed for observations of superconductivity,[25] the opening angle of the cells for diffraction was 30°, which limited the reciprocal space that was accessible and precluded full crystal structure refinement to determine atomic positions. Complete areal x-ray diffraction maps of the sample were measured at 124 GPa, and diffraction patterns were measured at selected pressures up to 178 GPa. The pressure was determined from diffraction of the Re gasket near the sample. A membrane drive was used to increase the pressure in the DACs. The x-ray diffraction images were collected using Pilatus detectors and integrated using the DIOPTAS software,[43] and the Rietica LHPM Rietveld program was used to analyze integrated powder diffraction patterns.[44]

**Acknowledgements**

We are grateful to A. Hameed, A. Mark, and S. A. Gramsch for useful discussions. This work was supported by the U.S. National Science Foundation (DMR-1933622, A.L. and R.H; and DMR-1809649, R.D.); the U.S. Department of Energy (DOE), National Nuclear Security Administration (NNSA), through the Chicago/DOE Alliance Center (CDAC; DE-NA0003975, R.K., M.A., and R.H.); and the DOE Office of Science, Fusion Energy Sciences (DE-SC0020340, N.S., R.D., and R.H.) X-ray diffraction measurements were performed at HPCAT (Sector 16) and GSECARS (Sector 13), Advanced Photon Source (APS), Argonne National

Laboratory (ANL). HPCAT operations are supported by the DOE-NNSA Office of Experimental Sciences. GSECARS is supported by the U.S. National Science Foundation (EAR-1634415). The Advanced Photon Source is a DOE Office of Science User Facility operated for the DOE Office of Science by ANL (DE- AC02-06CH11357).

**SUPPORTING MATERIAL**

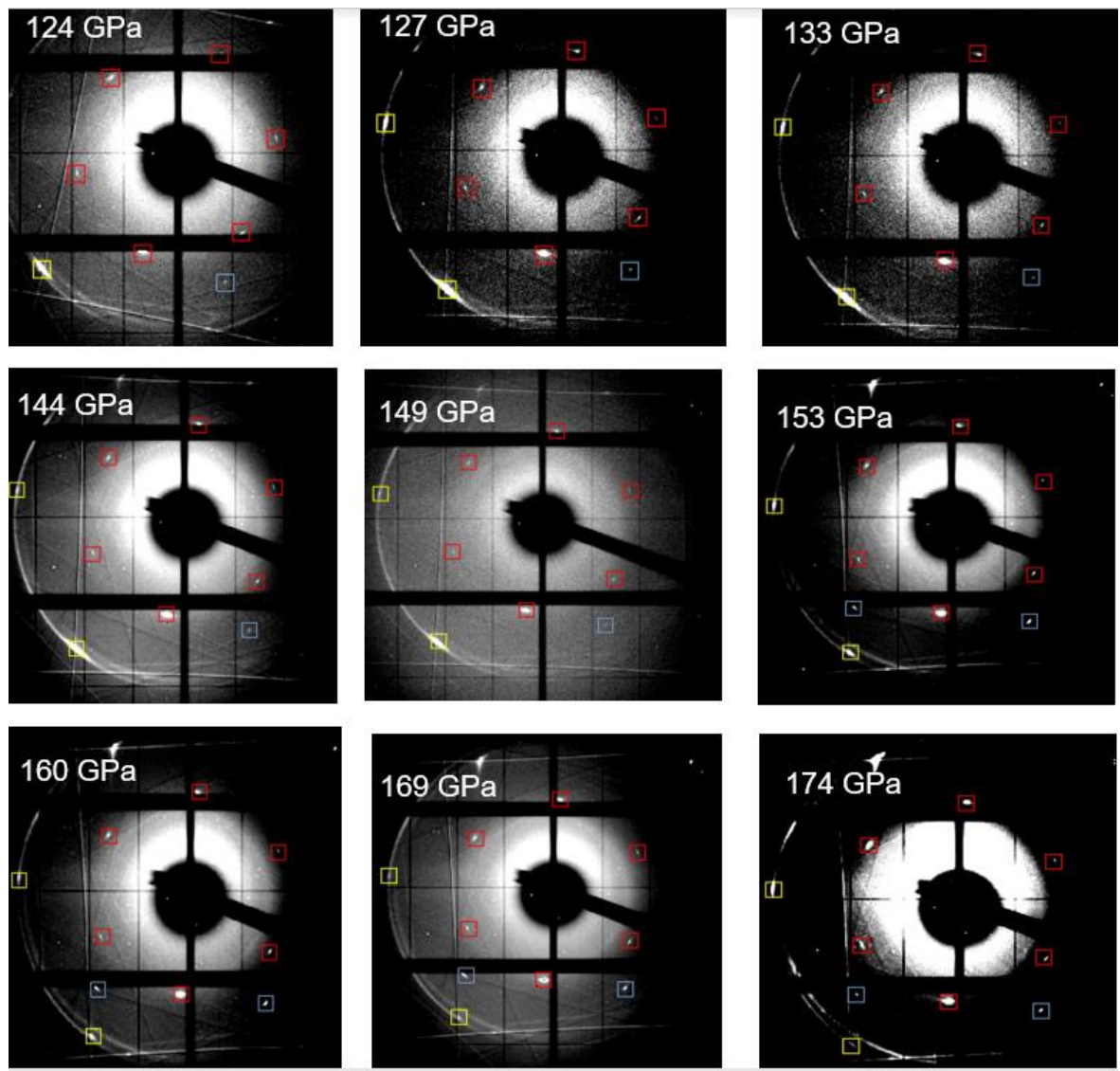

*Figure S1*. *Single-crystal diffraction images for the C-S-H sample obtained at HPCAT at various pressures. The orthorhombic indexing of the reflections is shown for 200/020 planes (red boxes), 201 planes (blue boxes) and 002 planes (yellow boxes). The black circle in the center of the image is the x-ray beam stop shadow and the black crosses arise from the detector.*

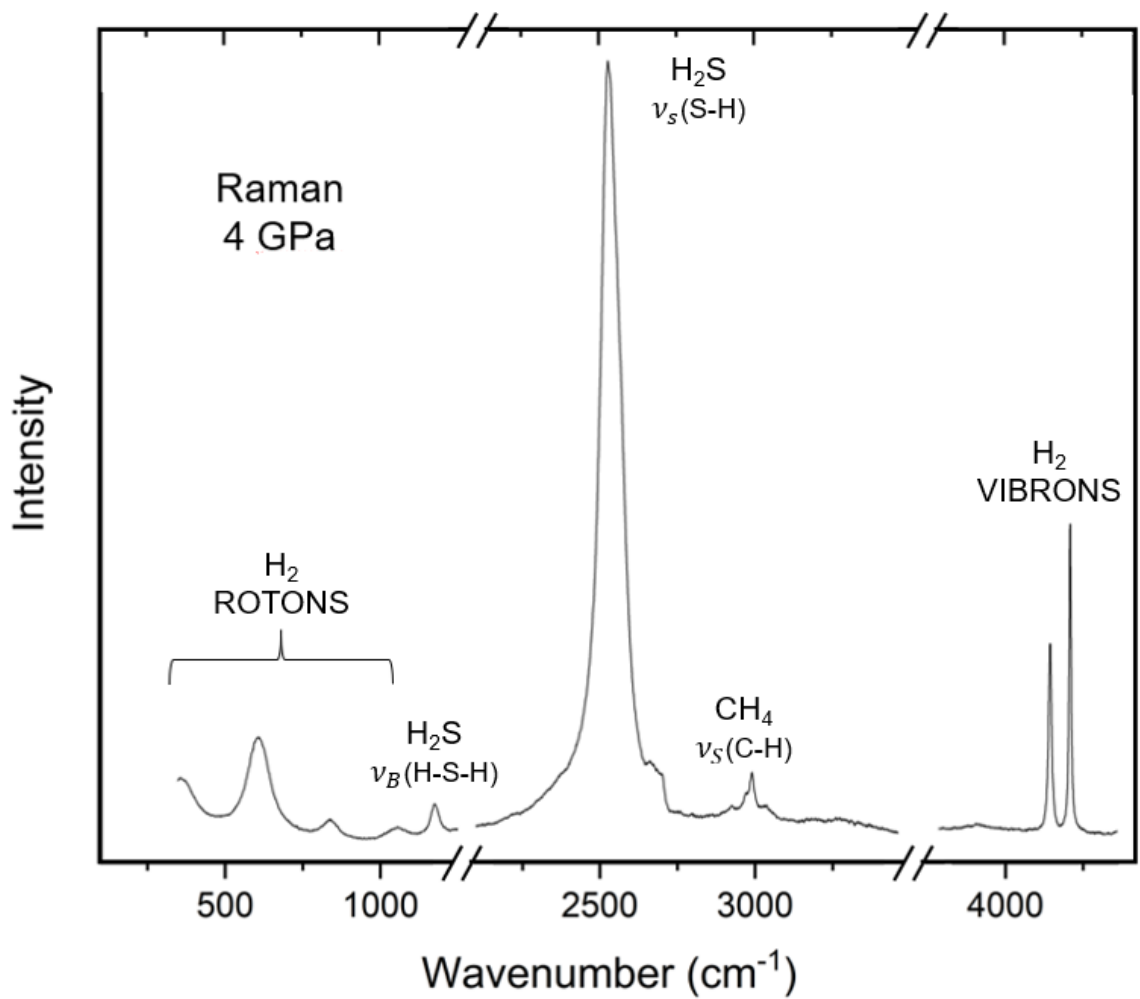

***Figure S2***. *Raman spectrum of a crystal of a $(H_2S)_{2-x}(CH_4)_xH_2$ van der Waals alloy synthesized from a $C+S+H_2$ mixture at 4 GPa.[25] Peaks corresponding to H-S-H bending, S-H stretching, C-H stretching, and $H_2$ rotons and vibrons are identified. Multiple peaks are observed in the region of the C-H stretching bands. $H_2$ vibrons split after the photochemically enabled synthesis. A H-C-H bending mode peak around 1500 $cm^{-1}$ is not shown.*